# SOLAR SYSTEM
# Voyager 2 enters interstellar space

*After 41 years of travel, the Voyager 2 spacecraft joins its twin in interstellar space. A suite of papers report Voyager 2's experience of its transition through the heliosheath and heliopause to what lies beyond.*

As the solar wind continuously blows away from the Sun at supersonic speeds, it creates a cavity in interstellar space filled with solar material; the heliosphere. Because the solar and interstellar plasmas have different compositions, densities, temperatures, and are braided by magnetic fields of different origin, they cannot interact freely and must be separated by a discontinuous boundary. This outer edge of the heliosphere is called the heliopause and marks the start of interstellar space, or rather, the start of the interstellar medium. The Voyager 1 spacecraft, launched in 1977 shortly after its partner Voyager 2, made the transition through the heliopause 7 years ago; now Voyager 2 has also completed this astonishing feat and data from several of its instruments taken during the crossing are reported in a series of papers in this issue[1–5].

The plasmatic influence of the Sun extends beyond the heliopause; the heliosphere, as an obstacle to interstellar inflow, can perturb the very local interstellar medium, perhaps even forming an interstellar bow shock, or wave, upstream of the heliopause. Of course, the gravitational influence of the Sun also extends well beyond the heliopause — at least as far as the Oort cloud — but, for the purposes of heliophysics, the heliopause represents the outer-most solar structure.

A complete and detailed understanding of the large-scale processes that shape the heliospheric structure, through its interaction with the interstellar medium, is not only important for heliophysics, but is also of interest to the astrophysics community; similar processes, most likely, also shape the structure of astrophysical jets, pulsar wind nebulae, jets, and other stellar wind collisions. The heliosphere also serves as a template for understanding the formation and dynamics of astrospheres around other stars, which can, in turn, have consequences for the habitability of planets housed within these protective bubbles. At present, only the heliosphere can be sampled in-situ, and only the Voyager spacecraft have sampled the outer heliospheric regions.

The Voyager 1 spacecraft crossed the heliopause in August 2012 at a radial distance of ~122 au from the Sun[6–8]. Although the charged particle intensities behaved as expected during the crossing (galactic cosmic ray intensities increased markedly, while the intensity of ions, accelerated within the heliosphere, decreased below detectable levels), magnetic field observations did not show a significant change in direction. This created confusion regarding this newly discovered boundary, which was initially termed the helio-cliff. Unfortunately, Voyager 1's plasma instrument was damaged in 1980, which made it impossible to directly identify the transition from the hot solar plasma to the colder and denser interstellar material beyond the heliopause. Confirmation of the heliopause crossing only occurred 8 months later when Voyager 1's plasma wave instrument recorded electron plasma oscillations[9], formed by locally shock-accelerated electrons streaming along magnetic fields, from which a plasma density consistent with interstellar values could be determined. The seemingly incongruous magnetic field observations across the heliopause were later explained by the draping of the interstellar magnetic field across the heliopause, forming a tangential discontinuity. Since then, Voyager 1 has been exploring deeper into interstellar space at rate of ~3 au yr$^{-1}$, with the hope of eventually sampling the unperturbed interstellar medium.

Unlike Voyager 1, Voyager 2 does have a working plasma experiment on board, and for this reason the Voyager 2 heliopause crossing was eagerly awaited. Results presented in this issue of *Nature Astronomy* confirm that Voyager 2 crossed into interstellar space on 5 November 2018 at a radial distance of 119 au[1-5]. This time, the heliopause crossing was confirmed by direct density measurements, and also shortly thereafter through electron plasma emissions. The Voyager 2 transition across the heliopause resembles that of Voyager 1, but with several significant differences: the second heliopause crossing was not pre-empted by a wide flow stagnation region while magnetic field measurements suggest a thinner and smoother heliopause structure with a stronger interstellar magnetic field directly beyond it. These detailed differences are yet to be explained, and may be partially due to temporal changes related to the Sun's dynamic solar cycle.

Even with the outer boundary of the heliosphere now sampled in-situ by two spacecraft, the structure of the unexplored regions remain a contentious subject; both Voyager spacecraft crossed the heliopause near the nose region (i.e. in the approximate direction of interstellar inflow), but there are no direct observations of the downwind tail region. And, as shown by ref. [10], the structure of the tail is determined by the strength of the interstellar magnetic field: if the interstellar plasma is magnetically dominated, the resulting heliopause should be spherical, while, if this is not the case, the heliopause will be comet-shaped, as is generally expected, and could extend for several thousand au in the downwind direction. Current Voyager observations close to the heliopause show relatively large magnetic field magnitudes, and, as argued by ref. [2], these could be sufficiently large to form a spherical and bubble-shaped heliosphere. The currently sampled magnetic field may also, however, be significantly compressed by the interaction with the heliosphere close to the heliopause, with the pristine field further out much weaker. This debate can only be settled by more direct observations from the Voyager spacecraft as they move further into interstellar space, exploring new regions and sending back new and unique data.

Unfortunately the Voyager mission cannot continue indefinitely and power requirements will force the mission to an end; perhaps within the coming decade. New strategies to extend the mission, including a new power management plan, were recently implemented and the real-time status of the spacecraft can be found at https://voyager.jpl.nasa.gov/. It is unclear when a follow-up mission will again reach interstellar space. Work is, however, ongoing in formulating such an interstellar probe in COSPAR's (Committee in Space Research's) Panel on Interstellar Research.

R. Du Toit Strauss
*Center for Space Research, North-West University, Potchefstroom, South Africa.*
*e-mail: dutoit.strauss@nwu.ac.za*

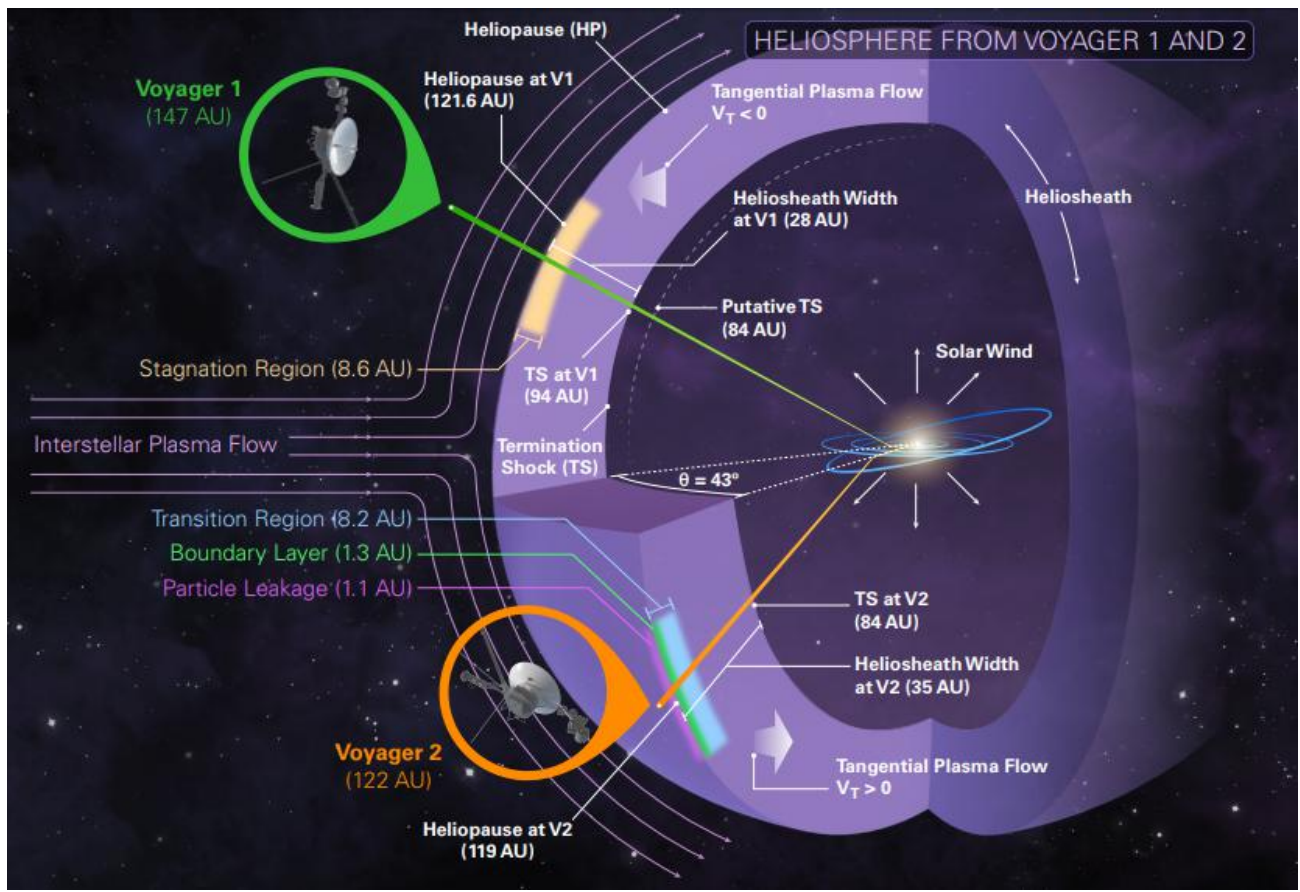

Fig. 1 | The heliosphere and very local interstellar medium with the structure of the heliopause emphasized (figure taken from ref. [2]).